\documentclass{ws-procs9x6}
\usepackage{graphicx}

\begin{document}

\title{General Relativity and Neutrino-driven Supernova Winds\footnote{\uppercase{T}his work was supported 
by  \uppercase{S}cientific \uppercase{D}iscovery \uppercase{T}hrough
\uppercase{A}dvanced \uppercase{C}omputing (\uppercase{S}ci\uppercase{DAC}), a program of the \uppercase{O}ffice of \uppercase{S}cience of the \uppercase{U.S}. \uppercase{D}epartment of \uppercase{E}nergy (\uppercase{DoE}); and by \uppercase{O}ak \uppercase{R}idge \uppercase{N}ational \uppercase{L}aboratory, managed by \uppercase{UT-B}attelle, \uppercase{LLC}, for the \uppercase{DoE} under contract \uppercase{DE-AC05-00OR22725}.}}

\author{C. Y. CARDALL}

\address{Physics Division \\
Oak Ridge National Laboratory \\
Oak Ridge, TN 37831-6354, USA \\
Email: cardallcy@ornl.gov}

\address{Department of Physics and Astronomy \\ 
University of Tennessee \\
	Knoxville, TN 37996-1200, USA}%

\maketitle

\abstracts{To date, detailed core-collapse supernova simulations have not succeeded in elucidating the explosion mechanism. But whatever the mechanism turns out to be, 
there will be a neutrino-heated outflow between
the hot, newly-born neutron star and the outgoing supernova shock wave whose neutron richness, low density, and high temperature could provide promising conditions for the synthesis of heavy nuclei via rapid neutron capture (the {\em r-}process). General relativistic effects improve the prospects for this outflow as an {\em r-}process site. Among relativistic effects, the enhanced ``gravitational potential'' is more important than the gravitational redshift or trajectory bending of neutrinos.}

\section{Core-collapse Supernovae}

Core-collapse supernovae---those 
of Type Ib, Ic, and II---result from the catastrophic collapse of the core of a massive star. For most of their existence, stars burn hydrogen into helium. 
In stars at least eight to ten times as massive as the Sun, 
temperatures and densities become sufficiently high to burn 
to carbon, oxygen, neon, magnesium, and silicon and iron group elements. 
The iron group nuclei are the most tightly bound, and here burning in 
the core ceases. 
The iron core---supported by electron 
degeneracy pressure---eventually 
becomes unstable. Its inner portion undergoes homologous collapse
(velocity proportional to radius), and the outer portion collapses 
supersonically. 
Electron capture on nuclei is one instability leading to collapse, and 
this process continues throughout collapse, producing neutrinos.
These neutrinos escape freely
until densities and temperatures in the collapsing core
become so high that even neutrinos are 
trapped. 

Collapse is halted soon after the matter exceeds nuclear density; 
at this point (called ``bounce''), a shock wave forms at the boundary between the homologous
and supersonically collapsing regions. The shock begins to move out,
but after the shock passes
some distance beyond the surface of the newly born neutron star, 
it stalls as energy 
is lost to neutrino emission and dissociation of heavy 
nuclei falling through the shock.

The details of how the stalled shock is revived
sufficiently to continue plowing through the outer layers of the
progenitor star are unclear. Some combination of neutrino heating of
material behind the shock, convection, instability of the spherical
accretion shock, rotation, and magnetic fields launches the explosion.

It is natural to consider neutrino heating as a mechanism for
shock revival, because neutrinos dominate the energetics of
the post-bounce evolution.
Initially, the nascent neutron star is a hot thermal bath of dense nuclear matter, 
electron/positron pairs, photons, and neutrinos, containing most of 
the gravitational potential energy released during core collapse. 
Neutrinos, having the weakest interactions, are the most efficient 
means of cooling; they diffuse outward on a time scale of seconds, 
and eventually escape with about 99\% of the released gravitational energy.

Because neutrinos dominate the energetics of the system, 
a detailed understanding of their evolution will be integral to any
detailed and definitive account of the supernova process.
If we want to understand the explosion---which accounts for only about 1\% of 
the energy budget of the system---we should carefully account for the
neutrinos' much larger contribution to the energy budget.

What sort of computation is needed to follow the neutrinos' evolution?
 Deep inside the newly-born neutron star, 
 the neutrinos and the fluid are tightly coupled (nearly in equilibrium);  but as the neutrinos are transported from inside the neutron star, they go from a nearly isotropic diffusive regime to a strongly forward-peaked free-streaming region. Heating of material behind the shock occurs precisely in 
this transition region, and modeling this process accurately requires tracking both the energy and angle dependence of the neutrino distribution functions at every point
in space. 

A full treatment of this six-dimensional neutrino radiation hydrodynamics
problem remains too costly for currently available computational resources. 
Throughout the 1990s, several groups performed simulations in two spatial dimensions with simplified neutrino transport.

One simplification allowed for neutrino transport in two 
spatial dimensions, but with
neutrino energy and angle dependence integrated out---effectively
reducing a five dimensional problem to a two dimensional one.\cite{hera94,burrows95,janka96}
These simulations showed convection in two regions. First, loss of electron neutrinos from the outer layers of the neutron star caused composition gradients that 
could drive convection, which boosted neutrino luminosities by bringing hotter material to the surface. Second, heating decreased further from the neutron star surface, giving rise to a negative entropy gradient. The resulting convection 
increased the efficiency of neutrino heating by delivering heated material to the region just behind the shock.
These simulations exhibited explosions, suggesting that the enhancements 
in neutrino heating behind the shock resulting from convection provided a
robust explosion mechanism. More recent simulations in three spatial
dimensions with this same
approximate treatment of neutrino transport showed similar outcomes.\cite{fryer02}

A different simplification of neutrino transport employed in the 1990s
was the imposition of energy-dependent
neutrino distributions from spherically symmetric simulations
onto fluid dynamics computations in two spatial dimensions.\cite{mezz98b} Unlike the multidimensional simulations
discussed above, these did not exhibit explosions, casting doubt
upon claims that convection-aided
neutrino heating constituted a robust explosion mechanism.

The nagging qualitative difference between multidimensional
simulations with different neutrino transport approximations 
renewed the motivation for simulations in which both the energy
and angle dependence of the neutrino distributions were retained.
Of necessity, the first such simulations were performed in
spherical symmetry (actually a three-dimensional problem, depending
on one space and two momentum space variables). Results from
three different groups are in accord: Spherically symmetric models
do not explode, even with solid neutrino transport.\cite{buras02,thompson02,liebendoerfer02}

Recently, one of these groups performed simulations in two spatial
dimensions, in which
their energy- and angle-dependent neutrino transport was made partially
dependent on spatial polar angle as well as radius.\cite{janka02,buras03}
Explosions were not seen in any of these simulations, except for one in which
certain terms in the neutrino transport equation corresponding to Doppler shifts
and angular aberration due to fluid motion were dropped. This was a surprising
qualitative difference induced by terms contributing what are typically thought of as small corrections. The continuing lesson
is that getting the details of the neutrino transport right makes a difference. 

Where, then, do simulations aiming at the explosion mechanism stand?
The above history suggests that elucidation of the mechanism will
require simulations that feature truly spatially multidimensional 
neutrino transport. Development of the formalism,\cite{cardall03} algorithms,\cite{cardall04} and
computer code necessary to this transport capability is
ongoing as part of the Terascale Supernova Initiative. This substantial 
collaboration---led by Anthony Mezzacappa of Oak Ridge National Laboratory and funded by the Department of Energy's Scientific Discovery through Advanced Computing (SciDAC) program---is dedicated
to the elucidation of the core-collapse supernova explosion mechanism through supercomputer simulations. At least one other major development of multidimensional neutrino transport capability is underway as well.\cite{burrows04}
In addition to sophisticated neutrino radiative transfer,
inclusion of magnetic field dynamics seems increasingly strongly motivated
as a possible driver of the explosion,
because simulations with ``better'' neutrino transport have failed to 
explode---even in multiple spatial dimensions. Work on the inclusion of magnetic fields is also part of the
Terascale Supernova Initiative.

\section{Relativistic Neutrino-driven Winds: Early Work}

It was realized in the early 1990s that whatever the explosion mechanism turns out to be, there will be an evacuated region between
the hot, newly-born neutron star and the outgoing supernova shock wave, whose neutron richness, low density, and high temperature of the region could provide promising conditions for the synthesis of heavy nuclei via rapid neutron capture (the {\em r-}process).\cite{meyer92}
Consideration of the exploding supernova models\footnote{These simulations were spherically symmetric. In light of the discussion in the previous section, it may be surprising that explosions were obtained. However, a prescription mocking up the multidimensional effects of a doubly-diffusive fluid instability (the so-called  ``neutron fingers'') was included in these simulations, which boosted neutrino luminosities sufficiently for the neutrino-driven explosion mechanism to succeed. That the necessary conditions actually exist for this particular instability to operate effectively has been disputed,\cite{bruenn95,bruenn96,bruenn04} but related instabilities may produce similar results.\cite{bruenn04}}  of Wilson and Mayle
in the late 1980s and early 1990s\cite{wilson93} led Meyer et al.\cite{meyer92}
to this insight, and to the related observation that the ejected amount of this
``hot bubble'' material---together with the Galactic supernova rate---seemed about
right to account for the amount of {\em r-}process material in the Galaxy. 

This proposal was buttressed by subsequent work of 
Woosley et al.:\cite{woosley94} The Wilson and Mayle models were
run out to several seconds past 
core bounce and explosion, showing that the intense neutrino fluxes
emitted by the cooling neutron star did indeed drive neutron-rich matter off its surface
into the evacuated region below the shock wave. This work also included nucleosynthesis 
calculations---performed by post-processing matter trajectories obtained in these simulations---that yielded impressive agreement with the observed solar system {\em r-}process abundance distribution.

The observed {\em r-}process abundances require that $\sim$100 neutrons 
be available for capture on each iron peak ``seed'' nucleus, and this 
neutron/seed ratio is determined by three parameters characterizing the
astrophysical environment: the entropy per baryon $S$, the electron fraction
$Y_e$, and the dynamic expansion time scale $\tau_{\rm dyn}$. High entropy
favors the relative disorder of free nucleons, as opposed to their being locked up in heavier nuclei (high rates of photodisintegration at the high temperatures and low densities associated with large $S$ provide the microscopic mechanism). Low electron fraction corresponds to neutron richness: $Y_e = n_p / (n_n + n_p)$, where
$n_p$ and $n_n$ are the number densities of protons and neutrons (including those locked up in nuclei). A short dynamic expansion time scale prevents too many seed nuclei from building up, by causing the freeze-out of bottleneck three-body reactions producing $^{12}$C from $^4$He.

The ``hot bubble'' exhibited in Woosley et al.\cite{woosley94} appears to be a subsonic outflow that bumps up against the shock wave sitting about 10$^4$ km from the neutron star.
In these conditions,  
the factor most favorable to the $r-$process turns out to be a high entropy of $S\sim 400$ (in units of Boltzmann's constant); values of $Y_e \sim 0.4$ and $\tau_{\rm dyn}\sim 1$ s are modest.

Several factors motivate modeling this neutron-rich outflow from the neutron star surface on its own---separate from large-scale supernova simulations---in order to
gauge its suitability as an {\em r-}process site. One obvious roadblock to the use of the large-scale simulations is the failure of most models to explode, as described in the previous section. Moreover, even if explosions are obtained, many of these simulation codes are not well suited to the task of running to the late times ($\gtrsim 10$ s) needed to follow the wind. A physical justification for simple wind models is that exploding two-dimensional supernova models with convection settled down to approximately spherically symmetric and stationary conditions as the wind phase approached.\cite{burrows95,janka96} Simple models can provide physical insight, and are more amenable to parameter studies (e.g. dependence on neutron star mass and radius, neutrino luminosities and average energies) than the large-scale simulations. 

Studies of simpler models of the neutrino-driven wind began with an important paper of  Qian and Woosley.\cite{qian96}
They obtained estimates of $S$, $\tau_{\rm dyn}$, and mass outflow rate from analytic ``wind'' models based on steady-state Newtonian fluid equations describing the matter outflow. They showed that the putative high entropy\cite{woosley94} of the ``hot bubble'' was difficult to explain, even with an outer boundary pressure. Their values of $S$ fell well short of that required for a robust {\em r-}process. They confirmed their analytic results with a hydrodynamic code that included simple input neutrino heating. In a few of their numerical runs, they employed an enhanced ``gravitational force'' $-{G M\over r^2}\rightarrow -{1\over 1-2G M/r}{G M\over r^2}$  motivated by the corresponding term in the relativistic fluid equations in Schwarzschild geometry. These cases resulted in larger values of $S$ and smaller values of $\tau_{\rm dyn}$, and it was pointed out that both went in the right direction towards more favorable conditions for the {\em r-}process.

\begin{figure}[t] 
\includegraphics[width=3.1in]{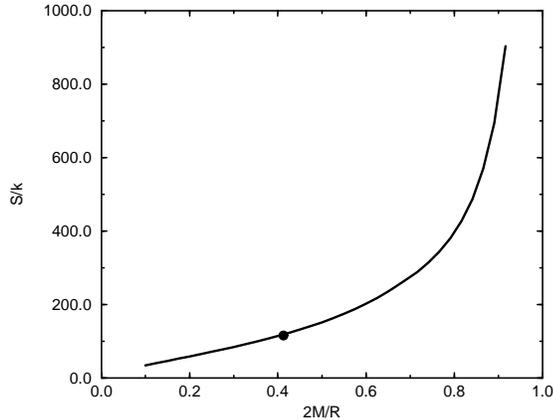}
\caption{The final entropy per 
	baryon in units of Boltzmann's
	constant, as a function 
	of the supernova core Schwarzschild radius
	divided by the core radius. 
	The circle is from the Qian and Woosley
	numerical calculation of 
	model 10B with post-Newtonian corrections. \label{entropyFigure}}
\end{figure}

\begin{figure}[t] 
\includegraphics[width=3in]{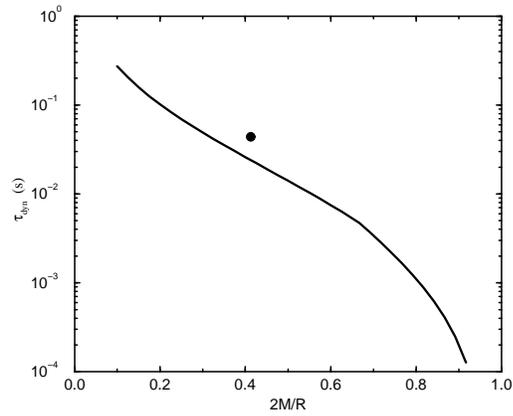}
\caption{The dynamic expansion time scale 
	as a function of the supernova core Schwarzschild radius
	divided by the core radius. 
	The circle is from the Qian and Woosley
	numerical calculation of 
	model 10B with post-Newtonian corrections. \label{expansionTimeFigure}}
\end{figure}

\begin{figure}[t] 
\includegraphics[width=3in]{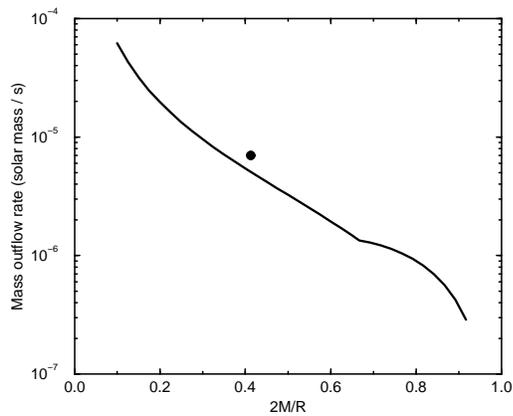}
\caption{The mass outflow rate as a 
	function of the supernova core Schwarzschild radius
	divided by the core radius. 
	The circle is from the Qian and Woosley
	numerical calculation of 
	model 10B with post-Newtonian corrections.\label{massOutflowFigure}}
\end{figure}

Cardall and Fuller\cite{cardall97} pursued this hint on the effects of
relativity, following the Qian and Woosley\cite{qian96} approach, but with
relativistic fluid equations. 
In addition to the relativistic effects in the fluid equations
(e.g. enhanced ``gravitational force,'' Lorentz factors limiting velocities to the speed of light, internal energy density and pressure contributing to inertia), relativistic effects on the neutrino heating were taken into account, specifically gravitational redshift and the bending of neutrino trajectories. 

These calculations affirmed that general relativistic effects make conditions in the wind more hospitable to the {\em r-}process. 
Figures 1, 2, and 3 show $S$, $\tau_{\rm dyn}$, and the mass outflow
rate as a function of the compactness of the neutron star (Schwarzschild radius divided by radius).\footnote{This variation was actually computed by varying the neutron radius with the mass held fixed at $1.4 M_\odot$. A particular prescription for the variation of neutrino luminosity was also employed.} Trends of increasing $S$ and decreasing $\tau_{\rm dyn}$ with increasing compactness are both favorable, though this comes at a cost of a smaller mass outflow rate, which translates into the production of less {\em r-}process material. These figures are not
physically meaningful for ${2G M\over R c^2} \gtrsim 0.66$, as compactness greater than this would imply an equation of state that violates causality (sound speed greater than the speed of light). Comparing Figures 1, 2, and 3 with a parameter study\cite{hoffman97} of the neutron/seed ratio as a function of $S$, $\tau_{\rm dyn}$, and $Y_e$, Cardall and Fuller\cite{cardall97} concluded that suitable 
conditions for the {\em r-}process could just be achieved near the neutron star
causality limit, but that the amount of ejected material becomes uncomfortably
(though perhaps not prohibitively) small.

The most important relativistic effect appears to be the enhanced  
``gravitational force.'' One way to see this is to note the good agreement
between the Cardall and Fuller\cite{cardall97} results and the single displayed 
Qian and Woosley\cite{qian96} numerical result, in which the enhanced
``gravitational force'' was the only relativistic effect included. This agreement
is particularly striking in the case of the entropy per baryon $S$. Roughly speaking, the entropy per baryon is the energy a baryon acquires by neutrino heating, divided by a characteristic  temperature at which the heating takes place. If a baryon is to escape the ``gravitational potential,'' the acquired energy per baryon is roughly the baryon mass times the gravitational potential at the neutron star surface.\cite{qian96} In the relativistic case, the ``gravitational potential'' obtained from the enhanced ``gravitational force'' expressed above is ${1\over 2}\ln\left(1-{2G M\over R c^2}\right)$, which reduces to the Newtonian expression ${G M \over R c^2}$ for small compactness. Gravitational neutrino redshift and trajectory bending have some effect on the characteristic temperature at which the heating takes place,\cite{cardall97} but this effect is modest in comparison with the logarithmic dependence of the ``gravitational potential.'' The result is that Figure 1 for $S$ tracks the logarithmic dependence of the gravitational potential quite closely; the corresponding plot for the Newtonian case would be a straight line matching the relativistic curve at low compactness. It is evident that relativity makes a nontrivial difference for neutron star masses and radii.

\section{Recent Work and Outstanding Issues}

Cardall and Fuller\cite{cardall97} concluded that the prospects for suitable {\em r-}process conditions in the neutrino-driven wind improve from something like `rather pessimistic' in the Newtonian case\cite{qian96} to perhaps `not inconceivable' when general relativity is taken into account---an assessment confirmed in subsequent
work.\cite{sumiyoshi00,otsuki00,wanajo01,thompson01} These works involved
various improvements on previous semi-analytic estimates, including full numerical solution of the wind equations; comprehensive variations of neutron star mass, radius, and neutrino luminosities; tracking of the electron fraction; construction of evolutionary sequences; proper treatment of transonic winds; and {\em r-}process network nucleosynthesis calculations. The consensus emerging from these works was that suitable {\em r-}process conditions might obtain, but that it would be in the form of a modest entropy, rapidly expanding (and possibly transonic) ``wind'' rather than a high entropy, subsonic ``hot bubble.'' The rather massive and compact neutron stars that seem to be required do not seem likely given current understanding of the dense nuclear matter equation of state, but new analyses of neutrino interactions with this dense matter may alleviate this problem by implying higher neutrino luminosities.\cite{reddy04}

A verdict of `not inconceivable' is not particularly comforting, but given the apparent preference of Galactic chemical evolution models for a supernova {\em r-}process source over neutron star binary mergers,\cite{argast04} fresh ideas on supernova winds are welcome. Hydrodynamic calculations have been performed in which outer boundary effects are claimed to play a significant role, allowing suitable {\em r-}process conditions for neutron stars of canonical mass;\cite{terasawa02} perhaps the issue of ``rapid wind'' vs. ``hot bubble'' is worth another examination. A very different and interesting idea is that magnetic fields may trap matter long enough to be significantly heated before being released into the wind.\cite{thompson03,thompson04}

Final understanding of the suitability of the supernova neutrino-driven wind as an {\em r-}process site will probably only come in the context of an understanding of the explosion mechanism---which brings us back to the discussion of large-scale simulations in the first section. Once this understanding is achieved through detailed simulation, we will have better qualitative and quantitative understandings of neutrino heating, magnetic fields, fallback of material at late times, and other phenomena that will influence the wind.

%
%
%
%

\def\prpts#1#2#3{{\it Phys. Reports} {\bf #1}, #2 (#3)}
\def\prl#1#2#3{{\it Phys. Rev. Lett.} {\bf #1}, #2 (#3)}
\def\prd#1#2#3{{\it Phys. Rev. D} {\bf #1}, #2 (#3)}
\def\prc#1#2#3{{\it Phys. Rev. C} {\bf #1}, #2 (#3)}
\def\plb#1#2#3{{\it Phys. Lett.} {\bf #1B}, #2 (#3)}
\def\npb#1#2#3{{\it Nucl. Phys.} {\bf B#1}, #2 (#3)}
\def\apj#1#2#3{{\it Astrophys. J.} {\bf #1}, #2 (#3)}
\def\apjl#1#2#3{{\it Astrophys. J. Lett.} {\bf #1}, #2 (#3)}
\def\apjs#1#2#3{{\it Astrophys. J. Supp.} {\bf #1}, #2 (#3)}
\def\aa#1#2#3{{\it Astron. Astrophys.} {\bf #1}, #2 (#3)}
\def\mnras#1#2#3{{\it Mon. Not. R. Astron. Soc.} {\bf #1}, #2 (#3)}

\end{document}